# Mixed-mode implementation of PETSc for scalable linear algebra on multi-core processors

Michèle Weiland, Lawrence Mitchell, Mark Parsons
EPCC
The University of Edinburgh
Edinburgh, UK
m.weiland@epcc.ed.ac.uk

Gerard Gorman, Stephan Kramer
Applied Modelling and Computation Group
Imperial College London
London, UK

*Abstract*— **With multi-core processors a ubiquitous building block of modern supercomputers, it is now past time to enable applications to embrace these developments in processor design. To achieve exascale performance, applications will need ways of exploiting the new levels of parallelism that are exposed in modern high-performance computers. A typical approach to this is to use shared-memory programming techniques to best exploit multi-core nodes along with inter-node message passing. In this paper, we describe the addition of OpenMP threaded functionality to the PETSc library. We highlight some issues that hinder good performance of threaded applications on modern processors and describe how to negate them. The OpenMP branch of PETSc was benchmarked using matrices extracted from Fluidity, a CFD application code, which uses the library as its linear solver engine. The overall performance of the mixed-mode implementation is shown to be superior to that of the pure-MPI version.**

I. INTRODUCTION

The Portable Extensible Toolkit for Scientific Computation (PETSc) [1][2][3] is a widely used library, maintained by Argonne National Laboratory, for the scalable solution of partial differential equations. PETSc offers a wide range of parallel data structures, linear and non-linear solvers as well as preconditioners. The implementations of the algorithms are optimised for performance and PETSc is often used as a key component of large scientific applications. One such application is the general-purpose, multi-phase computational fluid dynamics (CFD) code Fluidity [4]. Fluidity solves the Navier-Stokes equations and the accompanying field equations on arbitrarily unstructured finite-element meshes in up to three dimensions. It is used in areas including geophysical fluid dynamics, computational fluid dynamics, ocean modelling and mantle convection. Both Fluidity and PETSc are developed as open source projects that accept contributions from external developers.

Fluidity is currently being optimised for multi-core processors; the application is undergoing a transformation from being an MPI-only code to being parallelised in a mixed-mode fashion using OpenMP threading in addition to message passing. The two most computationally expensive parts of Fluidity are the finite-element global matrix assembly and the subsequent solving of the linear systems. PETSc is used as the main engine for the latter; it was decided that, rather than adding OpenMP at the application level it would be more beneficial to directly apply the mixed-mode programming approach to the library. This paper discusses both the implementation and the performance of the threaded PETSc library.

The work presented here was funded by the European Commission in FP7 as part of the APOS-EU project (grant agreement 277481) and Fujitsu Laboratories of Europe Ltd.

II. MOTIVATION

Over the past five years, processor architectures have evolved substantially. The trend towards more cores per processor, and less memory per core, is now firmly established. Using the UK's national HPC service (HECToR) [5] as an example, we can illustrate this evolution (see Table 1). When HECToR came into service in late 2007, the system used dual-core processors with a 2.8 GHz clock rate and 6 GBytes of memory per processor. Less than five years later, the number of cores per processor has increased by a factor of 8, the memory available per core has decreased by a factor of 3 and the processor clock rate has been lowered by 18%.

| *HECToR* | Q3 2007 | Q2 2009 | Q1 2011 | Q1 2012 |
| --- | --- | --- | --- | --- |
| Total cores | 11,328 | 22,656 | 44,544 | 90,112 |
| Cores per processor | 2 | 4 | 12 | 16 |
| Clock rate (GHz) | 2.8 | 2.3 | 2.1 | 2.3 |
| Memory per node (GB) | 6 | 8 | 16 | 16 |
| Memory per core (GB) | 3 | 2 | 1.3 | 1 |

Table 1: Changes in system configuration for HECToR since 2007.

*A. Hybrid Programming*

Multi-core processors are now ubiquitous in HPC and scientific codes need to be adapted to the specific characteristics of these processors if they are to fully exploit the current and future generation of HPC systems. Although the amount of memory available per core has reduced, the amount shared between cores has increased: a single shared-memory node on HECToR can address 32GB of memory. Programmers are effectively presented with two levels of parallelism: inside a node, cores share a contiguous memory address space and they can exchange information by directly manipulating this memory space; between nodes, an exchange of information is only possible through explicit communication. Exposing and expressing both intra- and inter-node parallelism can be achieved using a hybrid programming approach [6]. On the IBM BG/Q system for instance it was shown that the best performance for quantum-chromodynamics applications is achieved when using a single MPI process per node [7]. Fujitsu compilers for the K Computer and PRIMEHPC FX10 systems can exploit the VISIMPACT automatic parallelization technology that treats multiple cores on a node as a single CPU, automatically using threads in shared memory regions [8]. A popular model to achieve code hybridization is to combine OpenMP [9] threading (shared-memory, intra-node parallelism) with explicit message passing (distributed memory, inter-node parallelism).

*B. Reducing the Number of MPI Processes*

The motivation for moving away from MPI-only parallelised applications is not only given by the "fat node" architecture of multi-core based HPC systems, but also by the total size of these systems. Today's top supercomputers have hundreds of thousands of cores; this number will continue to go up over the coming years and applications are expected to eventually be able to use these increasing numbers of cores. Although work is being undertaken on MPI implementations to increase efficiency on large core counts [10][11], the costs of common MPI operations such as synchronization and global communication are likely to remain non-trivial. By adding shared memory threads, reducing the number of MPI processes concomitantly, we can alleviate some of these problems. There are also many algorithmic advantages relating to reducing the number of subdomains; for example scaling of multigrid methods is often constrained by the cost of coarsening across subdomains.

III. BENCHMARKING PLATFORM - AMD OPTERON 6200 "INTERLAGOS"

The main benchmarking system used for the work presented here is HECToR, a Cray XE6. The current configuration of HECToR is based on the AMD Opteron 6200 "Interlagos" processor series and Cray's Gemini interconnect [12]. The system's theoretical peak performance is 830 TFlop/s.

Interlagos is based on AMD's "Bulldozer" processor series [13]. An Interlagos processor has 16 cores, which are paired into modules, split over two dies. Figure 1 shows a schematic of a single module. Each core has its own L1 cache and integer scheduler; the L2 cache is shared across a module (2 cores) along with the floating point scheduler, which has two symmetrical "fused multiply-add" floating point pipelines. There are four modules per die; they share the 8MB L3 cache and have two links to 8GB of main memory. Four modules (or eight cores) thus make up one uniform memory access (UMA) region (see Figure 1). As a result, memory access even inside a processor is not uniform. A shared-memory node on HECToR consists of two processors (a total of 32 cores) and has four UMA regions.

Compared to previous processor generations, the architecture of the Interlagos multi-core processor is more involved and hierarchical. It is therefore an ideal platform to study the implications of the multi-core evolution on application performance and scalability, and a leading-edge test bed for the OpenMP enabled PETSc.

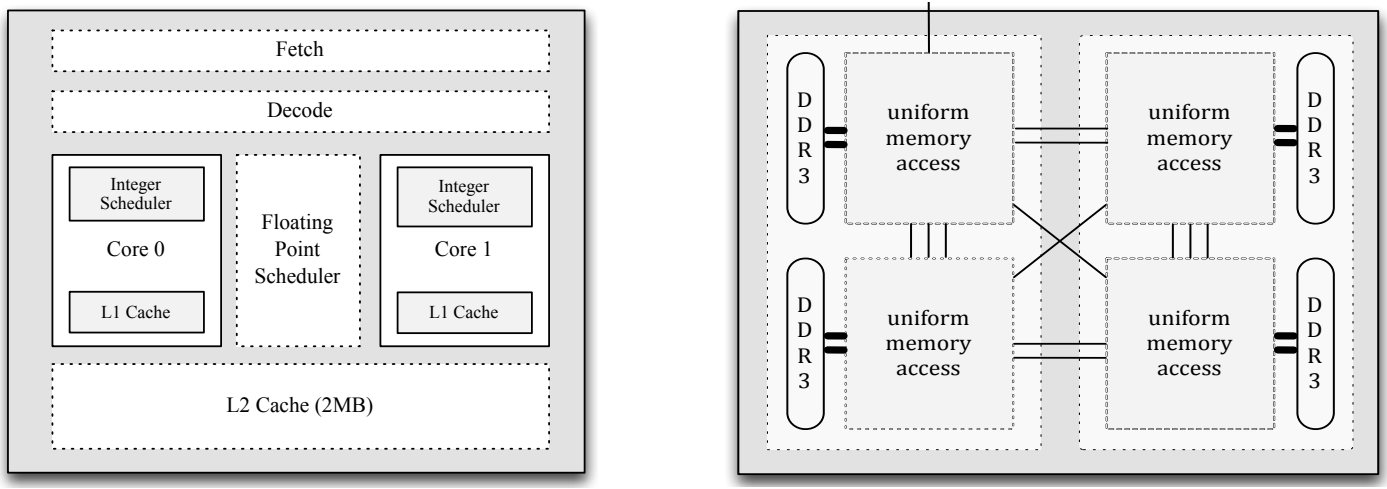

Figure 1: On the left, a schematic of an Interlagos "module, which contains a pair of processing cores. Two cores share a floating-point scheduler and L2 cache. The diagram on the right is the NUMA architecture of a Cray XE6 HECToR node. There are two UMA regions per processor, each connected to its own DDR3 memory bank.

IV. PERFORMANCE CONSIDERATIONS

Despite the mechanics of adding OpenMP directives to an existing code being a relatively simple process, achieving good performance from a threaded or hybrid application code can be difficult, especially on NUMA architectures. There are a number of performance-limiting factors that a developer needs to be aware of. We highlight the most common pitfalls and describe the solutions, which were also used in the process of threading and benchmarking PETSc.

*A. NUMA Effects and Memory Affinity*

The non-uniform memory access (NUMA) architecture was designed to overcome the scalability limits of the symmetric multiprocessing (SMP) architecture [14]. However this hierarchical memory model for multi-core processors means that it takes longer for a process or thread to access some parts of the memory than others. Processing cores inside an UMA region can directly read from and write to memory they are attached to via the memory bus. However if these cores need to access memory that is attached to another

UMA region on the same node, the read/write request needs to route via the HyperTransport links, increasing memory latency.

Optimizing data locality and managing memory affinity are therefore important issues to consider with multi-core processors that have a hierarchical memory model. Explicitly placing data into the memory of the processing core that will use these data will minimise memory latency and maximise cache reuse and memory bandwidth. The way the operating system manages memory can be used to one's advantage here: in Linux[1], page faults are by default bound to the memory directly connected to the CPU where the page fault was raised. Because the first CPU to touch the page will be the CPU that the page faults in, this memory management policy is called *first touch* [15]. More dynamic alternatives whereby pages could be migrated between memory nodes are also being investigated [16][17], but these are not currently part of the standard Linux kernel. The effects of memory affinity on a NUMA platform are straightforward to demonstrate, using the OpenMP threaded version of STREAM's Triad benchmark [18] as an example. Table 2 shows the performance with 32 OpenMP threads on HECToR, highlighting the performance benefits of exploiting the first-touch policy and faulting the memory in the correct pages by initializing the data in parallel. As a result the data is resident in core-local memory: memory bandwidth is increased and memory latency is minimised.

| *STREAM Triad (N=$10^9$)* | *Memory Bandwidth* | *Time* |
|---|---|---|
| Without parallel initialization | 21.80 GB/s | 1.10s |
| With parallel initialization | 43.49 GB/s | 0.55s |

Table 2: Running the STREAM Triad benchmark on a HECToR node using 32 OpenMP threads. Initializing the arrays in parallel, and thus optimizing data locality, improves the performance by a factor of two.

## B. Thread and Process Affinity

In addition to memory affinity, NUMA platforms also require the careful management of both thread and process affinity. Although memory affinity is dealt with inside an application, thread and process affinity need to be specified at the time of job execution. There exist a myriad of methods for specifying process affinity, each with a different interface: the standard Linux tool taskset [19]; Intel's thread affinity interface [20]; the GOMP CPU affinity environment variable [21]; or the platform-independent toolkit likwid [22]. On HECToR, the system's job scheduler provides this functionality. If threads and processes are not placed and pinned explicitly by the user, the OS will choose some default affinity. Furthermore, unless threads and processes are pinned to cores, the OS is free to migrate them. Both scenarios are likely to have a detrimental impact on performance and make it difficult to reproduce benchmarking results [23].

In addition to improved reproducibility of performance measurements, another major advantage of managing thread and process affinity explicitly is the opportunity to maximise the available memory bandwidth when either under-populating a node, i.e. using fewer cores than are available on the node, or running a hybrid application. Table 3 again shows the memory bandwidth achieved with STREAM's Triad, however this time the benchmark was executed with 4 OpenMP threads. The benchmark was submitted with threads pinned to cores explicitly. For the first two results, which are largely equal in performance, all the threads are placed on the same UMA region. However when placing the four threads across two or four UMA regions, the memory bandwidth increases accordingly. If it is necessary, for instance due to memory restrictions, to use fewer than the full set of cores on a node, placing the MPI processes or the OpenMP threads in an optimal fashion can improve performance considerably.

---

[1] Of all the systems that have participated in the June 2012 Top500, 462 list Linux as their operating system.

| STREAM Triad | Memory Bandwidth | Time |
|---|---|---|
| aprun –n 1 –N 1 -cc 0-3 ./stream | 6.64 GB/s | 3.78s |
| aprun –n 1 –N 1 -cc 0,2,4,6 ./stream | 6.34 GB/s | 3.79s |
| aprun –n 1 –N 1 -cc 0,4,8,12 ./stream | 12.16 GB/s | 1.97s |
| aprun –n 1 –N 1 -cc 0,8,16,24 ./stream | 30.42 GB/s | 0.79s |

Table 3: STREAM Triad benchmark on a HECToR node using 4 OpenMP threads, managing their placement explicitly.

## C. OpenMP Overheads

It is well understood that adding OpenMP threads to an application results in overheads related to the creation and management of the threads [24][25]. These overheads can have a significant impact on the overall performance of a threaded application. In the worst case, they may cancel out any performance benefits that threading offers. If the problem being computed inside a threaded region is small, the penalty incurred for creating a pool of threads can outweigh the benefits of dividing the computation up between them. It is therefore important to be aware how significant these overheads can be, especially since they can vary between parallel constructs, compilers and OpenMP runtime libraries. Table 4 shows the overheads that are incurred for the launch of a "parallel for" loop using different compilers on HECToR [26].

| *Overheads for* "parallel for" loop (μs) | *Number of OpenMP threads* | | | | | |
|---|---|---|---|---|---|---|
| | 1 | 2 | 4 | 8 | 16 | 32 |
| Cray 8.0.3 | 1.04 | 1.02 | 1.39 | 2.74 | 4.86 | 8.10 |
| GCC 4.6.2 | 0.55 | 1.16 | 5.94 | 21.65 | 50.15 | 88.40 |
| PGI 12.1 | 0.22 | 0.42 | 1.73 | 2.83 | 5.44 | 6.92 |

Table 4: OpenMP overheads (in μs) for the "parallel for" loop construct and the creation of a static loop schedule using the Cray, GCC and PGI compilers on HECToR.

## V. PETSc Code Structure

PETSc [1] consists of a suite of libraries that implement the high-level components required for linear algebra in separate classes: Index Sets, Vectors and Matrices; Krylov Subspace Methods and Pre-conditioners; and Non-linear Solvers and Time Steppers. The structure of the suite led our decisions regarding what parts of the code required threading.

### A. Vec and Mat Classes

The Vector and Matrix classes represent the lowest level of abstraction and are the core building blocks of most of the functionality in the suite. PETSc offers two Vector types: sequential and parallel. The implementation of the parallel vector type uses the sequential type, distributed over a specified number of MPI processes. The Matrix class is analogous to the Vector class – the matrix rows are distributed among the MPI processes, where each row is split into a "diagonal" part that contains the local column entries, and an "off-diagonal" part that contains the column-entries associated with indices assigned to other processes. These rows are then locally stored in a diagonal and off-diagonal matrix for which the sequential Matrix class is used.

PETSc has support for compressed row sparse storage (CSR, the default type), dense storage and block storage. Within PETSc, the CSR format is also referred to as AIJ. A positive side effect of the design of the suite is that by threading the sequential functionality, the parallel classes essentially pick this threading up for free. For example, a threaded operation on a sequential vector will also be available to a parallel vector, which itself is made up from sequential vectors. There are a handful of exceptions: most notably the initialization of parallel vectors, which, although they use sequential functions on the local parts, are set up in a separate function.

*B. KSP and PC Classes*

The Krylov subspace methods and the pre-conditioners are implemented in the KSP and PC classes. It was established that no threading was required in the KSP class. This is because nearly all the computation in methods such as Conjugate Gradient (CG) or Generalised Minimal Residual (GMRES) is concentrated within basic vector operations and sparse matrix-vector multiplications. These are already threaded in the Mat and Vec classes, and thus methods in the KSP class will use them automatically. Simple pre-conditioners (such as Jacobi) are based on functionality from the Mat and Vec classes that are threaded. It is also possible to build and link PETSc with multi-threaded third-party pre-conditioners such as Hypre [27]. Additionally, a geometric/algebraic multigrid framework (PCGAMG) that uses Chebyshev smoothers is in development in PETSc, the main components of which again consist of the already threaded Mat and Vec methods. However, the threading of many other frequently used preconditioners, such as Symmetric Over-Relaxation (SOR) or Incomplete LU-decomposition (ILU), is difficult due to their complex data dependencies and may require a redesign of the algorithms. This therefore falls without the scope of this work and, given the alternatives mentioned above, it was decided not to multi-thread any functionality in the PC class directly at this point in time.

*C. Existing pthreads Implementation*

A small number of matrix and vector operations in version 3.2 of PETSc have been threaded using pthreads [28]. The reason for investing effort in implementing an OpenMP threaded version of PETSc as well is that of improved interoperability with user applications. If a user code is hybridised with OpenMP threads, but is required to use pthreads inside PETSc, two thread pools will be maintained (one for OpenMP and one for pthreads) at all times, resulting in an over-subscription of the available resources. Offering an OpenMP threaded implementation of the PETSc library means that a single thread pool can be used across both the application and the library.

VI. IMPLEMENTATION

Figure 2 shows all the vector and matrix operations that were threaded as part of this effort to multi-thread some of the main functionality inside PETSc. Two main concerns guided the implementation choices: firstly, the responsibility for the performance of the threaded library should, as far as possible, not be with the user; and secondly, NUMA-awareness should be ensured through correct memory management.

*A. Data Memory Paging*

Two options present themselves with regards to the implementation of the library's memory paging behaviour. The first option is for PETSc to never page any data memory for threaded objects, offloading this concern entirely to the user application. The second option is for PETSc to take complete responsibility for managing the memory layout for threaded objects. In the first case, the user application needs to ensure that all the data are paged correctly in order to achieve good memory performance. As one of the design criteria dictates that the library's performance should not rely on user choices, this avenue is not pursued. In the second case where PETSc manages the memory layout, we have a choice of mapping the data either statically or dynamically. For the former, the same static schedules are used throughout the code. For the latter, data layout entries are added to all threaded objects, which need to be queried by the user code in order to find the correct OpenMP allocation ranges.

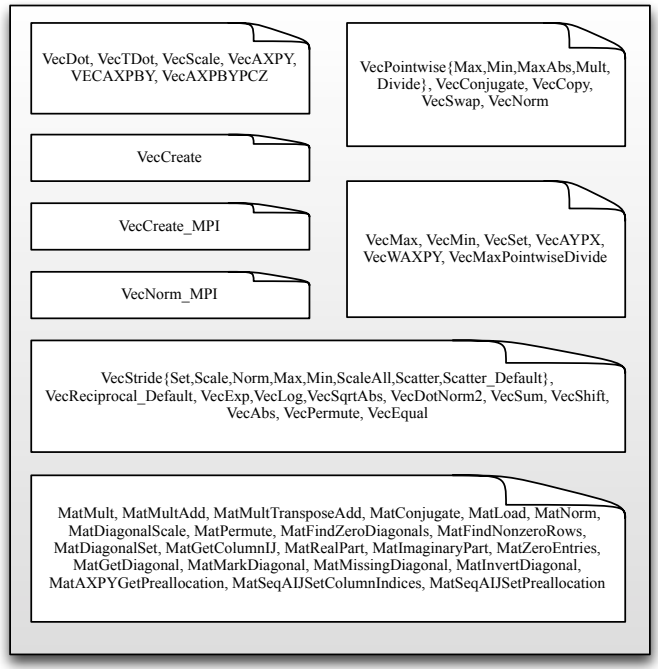

Figure 2: List of functions in the Vec and Mat classes that have been threaded with OpenMP.

Using a dynamic memory mapping approach would require the user application to correctly query all threaded objects, introducing an additional level of complexity to the code. However, PETSc "zeros" (i.e. touches) memory for all allocated vectors, as well as for matrices for which pre-allocation has been set up. This feature can be used to page memory correctly by using the same static schedule for all objects. Our implementation ensures that PETSc pages all threaded objects using an OpenMP static schedule and that the user application is only required to map the data using the same static schedule for good memory performance.

Matrix-vector multiplication operations are the basis of Krylov subspace solves and their performance is often limited by memory bandwidth; we therefore decided that the memory paging should favour this operation specifically. Knowing that each matrix row is multiplied by the vector, we page the matrix data by rows as shown in Figure 3. Threads now hold complete rows of the matrix in their local memory and can perform the multiplication keeping remote memory access to an absolute minimum.

*B. Level 1 BLAS*

Some of the vector operations are based on calls to Level 1 BLAS functions. In most BLAS implementations (Intel's Math Kernel Library being a notable exception [29]), these functions are not threaded. Calling them directly from within a multi-threaded application therefore results in a serial bottleneck. The solution implemented for PETSc is to parallelise calls to BLAS functions on the library level by calling the functions for a portion of a vector on each thread. Level 1 BLAS functions are called inside parallel regions; each thread calculates the boundaries of its portion of a vector depending on the number of threads available and the ID of the calling thread.

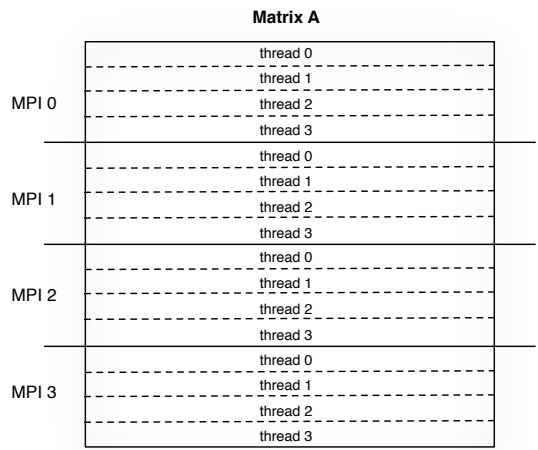

Figure 3: Diagram of dividing a matrix up between 4 MPI processes, each with 4 OpenMP threads, and paging the memory by rows.

## C. Generic Parallel Sections with Macros

Many of the OpenMP parallel sections in PETSc follow a generic pattern. It was decided to replace the OpenMP pragma statements with macros that indicate the start and end of an OpenMP parallel region. The macros translate into the OpenMP code if threading is enabled, otherwise revert to an implementation without OpenMP. Table 5 shows how the macros are used in a Vector function to specify a *parallel for* loop and its arguments in a generic fashion (lines 9 and 12).

*VecConjugate_Seq in src/vec/vec/impls/seq/bvec2.c*

```
1 PetscErrorCode VecConjugate_Seq(Vec xin)
2 {
3   PetscScalar    *x;
4   PetscInt       i;
5   PetscErrorCode ierr;
6
7   PetscFunctionBegin;
8   ierr = VecGetArray(xin,&x);CHKERRQ(ierr);
9   VecOMPParallelBegin(xin, default(none) private(i) shared(x));
10    for (i=__start; i<__end; i++)
11      x[i] = PetscConj(x[i]);
12  VecOMPParallelEnd();
13  ierr = VecRestoreArray(xin,&x);CHKERRQ(ierr);
14  PetscFunctionReturn(0);
15 }
```

Table 5: Example of an OpenMP pragma statement (a parallel for loop in this case) with generic macros.

An advantage that macros can bring is the ability to switch the multi-threaded parallelism on or off, depending on the size of the objects that are being used. As mentioned earlier, the overheads of OpenMP may negate any performance benefits for small data, thus being able to individually switch off OpenMP parallel regions for small objects can bring a performance benefit. The function that, based on object size, decides whether or not to switch off threading can be implemented at a high level as part of the generic macros, thus keeping the core library implementation unchanged.

## VII. Performance Factors for Sparse Matrix-Vector Multiplication

PETSc represents distributed MPI matrices by dividing them into on-diagonal and off-diagonal parts, which on each process are simply stored as sequential matrices (see Figure 4 for an illustration of a matrix that is distributed over eight MPI processes). As a consequence of this storage strategy, the matrix-vector multiplication is implemented as follows: on each process, the on-diagonal matrix elements are multiplied with their corresponding local vector elements; the vector elements that reside off-process are scattered into the a sequential vector in the local memory of the executing process; finally the off-diagonal matrix elements are multiplied with the formerly remote vector element which are now also local to the executing process, and the two partial solutions are added together.

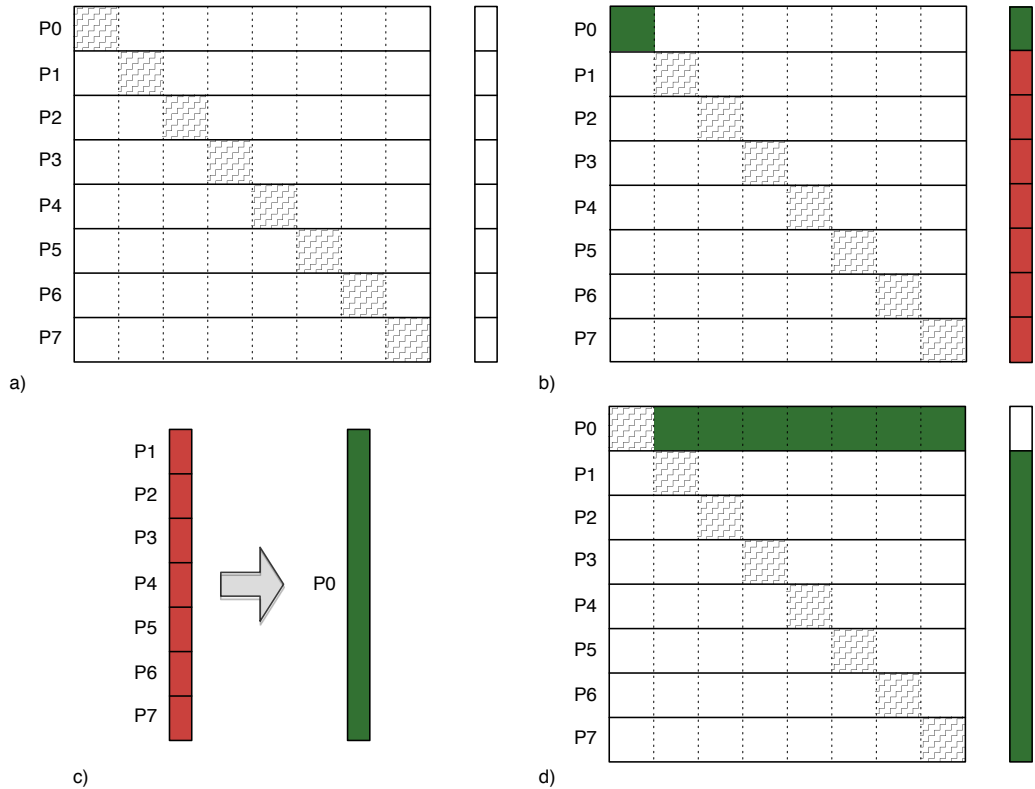

Figure 4: a) Distribution of a matrix and a vector across 8 MPI processes. PETSc distinguishes between on-diagonal and off-diagonal matrix entries. b) The on-diagonal matrix entries are multiplied with the local portion of the vector (in green). c) The remote vector entries (in red) are scattered to the executing process. d) The off-diagonal matrix entries are multiplied with the remainder of the vector (in green), whose entries are now local to the executing process.

The hybrid version of the matrix-vector multiplication builds on the MPI-only implementation; the distinction between on-diagonal and off-diagonal matrix elements remains, however the data that resides on each process is distributed among different threads (see Figure 5 a for an illustration of a matrix distributed over 4 processes with 2 threads per process). Similarly to MPI-only, the on-diagonal elements are multiplied first; each thread is responsible for its portion of those elements. Both matrices and vectors are paged by row and as a result, only part of the vector that resides on the executing process will have good data locality with thread that is computing part of the on-diagonal product. Figure 5 gives an example: the on-diagonal part of the matrix on process 0 is shared between 2 threads. Each thread will multiply its local matrix elements with the vector elements on process 0; elements local to each thread are coloured the same. Thread 0 will need to access vector elements that are local to thread 1 and vice versa. The same is true for the off-diagonal elements: the remote vector elements are scattered to the executing process, where they are stored in a sequential vector. This vector is again paged by rows, which means that threads will need to access shared data that is not local to them.

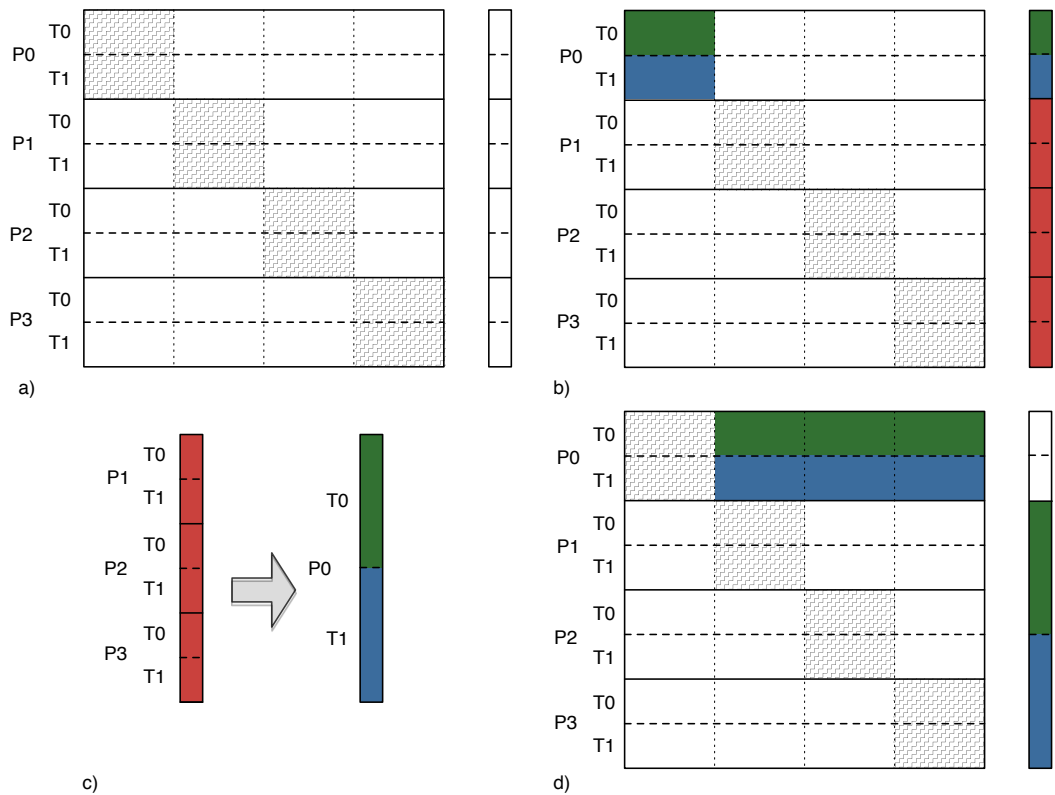

Figure 5: a) Matrix distributed among 4 MPI processes with 2 threads each. b) The threads are responsible for their portion of on-diagonal elements and need to access some vector elements that are shared, but non-local. Here, the elements that are local to thread 0 are coloured in green, those that are local to thread 1 are blue. c) The remote vector elements (in red) are scattered to the executing process and stored in a sequential vector, which is paged by row. d) Similarly to the on-diagonal elements, threads are responsible for their off-diagonal portion of the matrix.

Different factors influence the performance of both the MPI-only and the hybrid implementation of the matrix-vector multiplication. The scattering of the vector elements and the initial on-diagonal multiplication are allowed to overlap, however the collective MPI communications involved in the scattering step can be a performance bottleneck for large numbers of processes. For equivalent numbers of cores, the hybrid version has an advantage here: a lower number of MPI processes means that the amount of data in the on-diagonal parts is larger and, as a consequence, less data needs to be gathered from remote processes and communication bottlenecks are reduced. As a side effect, load imbalance will be improved. The hybrid version however suffers in terms of vector data locality, as threads need to repeatedly access data that is not local to them; memory access speeds and cache reuse will be reduced. If the threads span across more than one NUMA region, the effect will become more pronounced.

Although good data locality was one of our main concerns when implementing a threaded version of PETSc, the choice of paging vectors by row can bring some performance penalties. A solution would be to ensure that each region of uniform memory access has it own complete copy of the vector, sacrificing free memory for access speed. This solution involves more significant modifications of the PETSc data structures (making them "hybrid" aware) and will be investigated in the future.

## VIII. BENCHMARKING AND PERFORMANCE ANALYSIS

The implementation of the OpenMP threaded functionality in PETSc was tested in two ways:

- on a single-node level, directly comparing the performance of pure MPI to OpenMP;
- across nodes, assessing the hybrid MPI/OpenMP behaviour.

As the motivation for the work originated from the desire to obtain a hybrid version of the Fluidity application, benchmarking focused on solving linear systems extracted from problem cases typical to CFD.

### A. Tests Matrices and Code

The Fluidity team has developed a suite of HPC benchmarks, which is used for profiling and scalability testing. These tests are derived from the existing development testing framework [30]. The tests cover a wide range of laboratory-scale to ocean-scale scenarios using a variety of numerical schemes applicable to these problems as used in scientific studies. The matrices used in these tests were extracted and used in a standalone PETSc benchmark. The matrices represent the following benchmarks:

- *Backward Facing Step*: The backward facing step is a classical 3D fluid problem in the study of turbulent flows. It is widely used as a benchmark to evaluate the performance of turbulence models in separating flows.

- *Lock Exchange*: The lock exchange is another classic CFD test problem. A lock separates two fluids of different densities (e.g. hot and cold) inside a tank; when the lock is removed, two gravity currents propagate along the tank.

- *Salt fingering*: This matrix is assembled in the modelling of a 2D process in which warm salty water flows over colder, non-salty water. The fluids will mix: the warmer water sinks and loses both its heat and salt, while the displaced colder water starts rising up, gaining in temperature and salt levels.

- *Flue*: Fluidity is used to create a high fidelity model of the dispersion of a momentum driven flue emission in a cross wind under neutral atmospheric conditions. The matrix used here is for pressure, which is solved separately from velocity using the projection method.

Table 6 lists the sizes of the test matrices, together with the number of non-zero elements (NNZ); all the matrices listed in this table were used to confirm our performance results. The benchmark code that was used to test the performance of the threaded PETSc functionality (ex6.c) is taken from the suite of KSP examples that are part of the library; it is a generic benchmark that reads a PETSc matrix and vector from a file and solves a linear system. The problem definition is highly customizable and users can specify the solver and pre-conditioner, the maximum number of iterations and the convergence tolerances, to name but a few options. A single test can thus be used to benchmark the performance of a range of functions

### B. Renumbering the Matrices

It is generally difficult to achieve good cache reuse for sparse matrices. Typically, the benefits of matrix reordering are only realised when the problem is sufficiently small such that it can be largely contained within cache. However, architecture features such as shared cache between cores help with this. For the performance analyses presented here, the Reverse Cuthill-McKee (RCM) algorithm [31] was used on the test matrices to minimise their bandwidth.

Figure 6 shows the sparsity patterns of the "Backward Facing Step" velocity matrix before and after reordering using the RCM algorithm. For performance reasons the nodes and edges of the mesh are in practice reordered by the application code prior to any matrix assembly and solve; we reordered all our test matrices separately before using them as benchmarking data.

| Test Case | Matrix | Rows | Columns | NNZ |
|---|---|---|---|---|
| Lock-Exchange | Pressure | 64,750 | 64,750 | 4,337,952 |
| Backward Facing Step | Pressure | 263,477 | 263,477 | 18,642,163 |
| | Velocity | 790,431 | 790,431 | 11,294,379 |
| Saltfingering | Temperature | 688,086 | 688,086 | 14,112,698 |
| | Velocity | 1,376,172 | 1,376,172 | 9,632,240 |
| | Pressure | 688,086 | 688,086 | 14,112,674 |
| | Geostrophic pressure | 688,086 | 688,086 | 4,816,114 |
| Flue | Pressure | 10,079,144 | 10,079,144 | 747,090,670 |

Table 6: Sizes and numbers of non-zero elements for each of the test matrices.

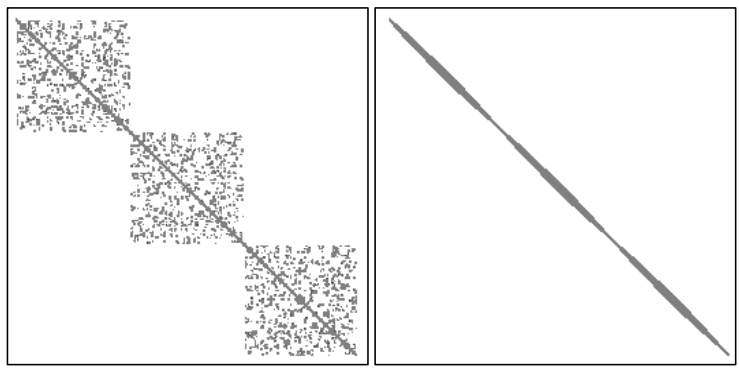

Figure 6: "Backward Facing Step" velocity matrix: on the left is the matrix's original sparsity pattern, on the right is the sparsity pattern after applying the RCM reordering algorithm.

## C. Single Node Performance

Good hybrid performance across many nodes can only be expected if the implementation manages the NUMA architecture correctly and exploits the advantages that a multi-core processor can offer. Single node threaded performance was therefore tested thoroughly; aside from pure performance, a few other aspects were tested as well:

- the difference in performance between executables built with and without OpenMP enabled;
- the impact of compilers on the scalability and runtime;
- the effect of explicit process and thread pinning when under-populating a node.

### 1) Impact of Compiler Choice

On HECToR the threaded PETSc library and the benchmark code were built with two different compilers, Cray 8.0.3 and GNU 4.6.2, allowing us to verify that their performance is not dependent on the choice of compiler. In addition, the library and the benchmark were built once with OpenMP disabled, and once with OpenMP enabled. When OpenMP is disabled, the benchmark is truly MPI-only and the compiler ignores all OpenMP statements in the code. Figure 7 shows the impact of both these aspects for the MatMult function, which performs the sparse matrix-vector multiplication, and is here part of a GMRES solve on the

Saltfingering Geostrophic Pressure matrix[2]. The plot on the left compares the runtime of the "pure" MPI benchmark with one that was compiled with OpenMP support and executed with OMP_NUM_THREADS set to 1. The version that was compiled with OpenMP enabled is marginally faster than the pure MPI version for the smallest core counts. An explanation for this is most likely improved compiler optimization: the compiler can use the additional information contained in the OpenMP statements, such as which variables are private, to make optimization decisions that benefit performance. If OpenMP is disabled at runtime, this performance benefit is not lost, as the overhead of the OpenMP runtime is negligible when only one thread is used. The Cray compiler seems more capable of this optimization than the GNU compiler, presumably due to its more advanced analysis. The right-hand plot shows runtimes for the same MatMult, this time comparing the OpenMP versions with the Cray and the GNU compilers. The runtimes achieved with the two compilers differ due to different costs of overheads that can be associated with the OpenMP runtime. The difference in performance observed here is almost negligible. The two graphs can also be compared to each other directly as they both show the runtimes for the same MatMult; for both compilers, it can be seen that the threaded code outperforms the MPI code on all core counts.

| KSPSolve Saltfingering Pressure | 1 | 2 | 4 | 8 | 16 | 32 | | MatMult Saltfingering GeoPressure | 1 | 2 | 4 | 8 | 16 | 32 |
|---|---|---|---|---|---|---|---|---|---|---|---|---|---|---|
| MPI with craycc | 96.640 | 55.389 | 44.029 | 38.374 | 19.489 | 11.139 | | Pure MPI with craycc | 46.790 | 26.052 | 17.919 | 15.116 | 8.151 | 4.388 |
| OpenMP with craycc | 96.640 | 67.503 | 41.271 | 33.967 | 16.768 | 9.134 | | MPI with craycc | 46.581 | 30.482 | 17.257 | 15.429 | 8.073 | 4.369 |
| MPI with gcc | 104.080 | 60.145 | 44.226 | 38.425 | 19.471 | 11.112 | | OpenMP with craycc | 46.581 | 27.107 | 15.152 | 11.834 | 5.580 | 2.689 |
| OpenMP with gcc | 104.080 | 66.618 | 40.603 | 34.322 | 16.601 | 9.368 | | MPI with gcc | 52.154 | 35.076 | 18.931 | 15.829 | 8.000 | 4.309 |
| | | | | | | | | OpenMP with gcc | 46.636 | 28.780 | 15.219 | 12.373 | 5.751 | 2.830 |
| | | | | | | | | Pure MPI with gcc | 52.154 | 35.076 | 18.931 | 15.829 | 8.000 | 4.309 |
| | | | | | | | | MPI with gcc | 46.636 | 31.583 | 17.723 | 15.756 | 8.076 | 4.295 |

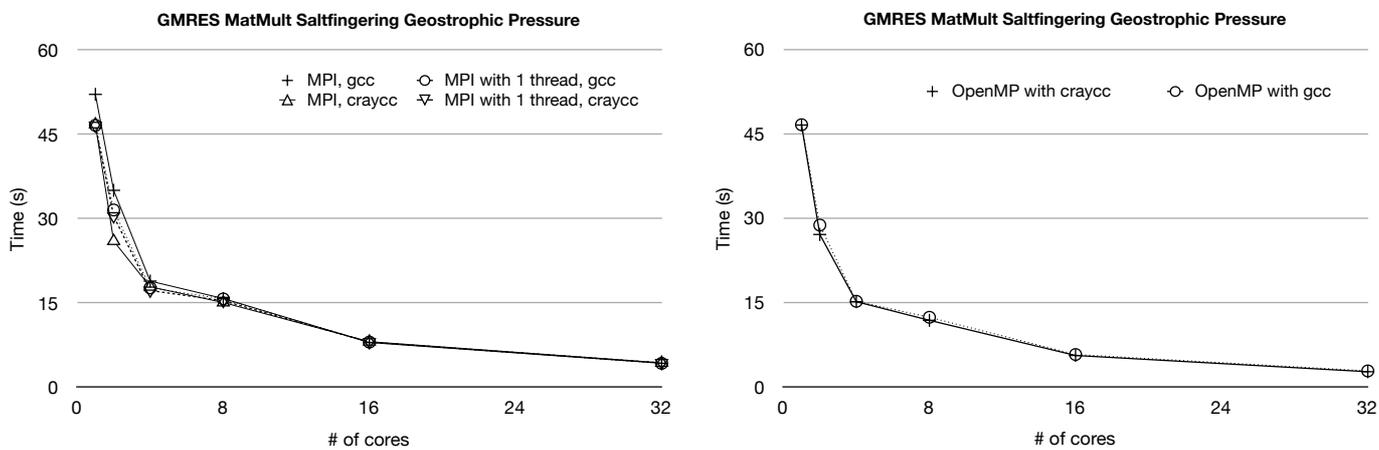

Figure 7: Two graphs that plot the time spent inside sparse matrix-vector multiplications during a GMRES solve of the Saltfingering Geostrophic Pressure matrix. The graph on the left compares the runtimes for pure MPI with an MPI version that was built with OpenMP enabled for the GNU and Cray compilers. The graph on the right compares the pure OpenMP performance.

*2) Default versus Explicit Affinity*

Cray's Application Level Placement Scheduler (ALPS) [32] provides users with process affinity control through its job launch command *aprun*. The *–cc cpu_list* option is used to define explicitly what cores are to be used during the execution of a job. This is mainly relevant when under-populating a node, in this case whenever fewer than 32 cores per node are being used.

Figure 8 shows how explicitly controlling process and thread affinity can impact performance, and thus scalability, of a code. The graph on the left compares the scalability of the MatMult-part of a CG solve on the Backward Facing Step Velocity matrix using default and explicit process affinity. Looking at the default affinity, both the MPI and the OpenMP scaling curves are similar, with strong scaling dropping off at 4 cores and a parallel efficiency of around 50%. When pinning processes and threads explicitly, the scaling behaviour of MPI and OpenMP is again similar, however compared to the default behaviour the scalability is much improved.

---

[2] Unless stated otherwise, performance results presented in this paper were achieved on HECToR and are as reported by PETSc's internal log functionality.

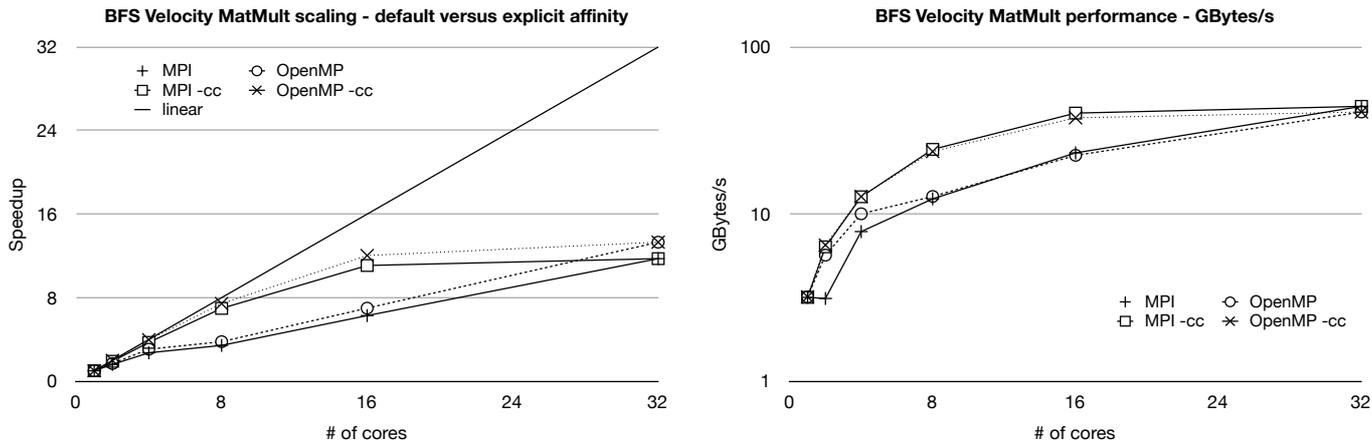

Figure 8: The scaling behaviour of the MatMult component of a CG solve on the Backward Facing Step Velocity matrix with both default process/thread affinity and explicit process/thread pinning (achieved by using the "-cc" option with the aprun job launcher) is shown on the left. The graph on the right shows the corresponding memory bandwidth, as reported by CrayPat.

We experimented with different process placements for every core count and only plotted the best performance. The OpenMP code achieves approximately 75% parallel efficiency on 16 cores (this is slightly less for the MPI code at 70% parallel efficiency). The reason for this improvement is the increase in memory bandwidth: when using the default placement, processes are assigned to the cores in a round-robin fashion and processes will by default be placed closely together, thus creating contention on the memory infrastructure. If the processes are placed further apart however, they will be able to make use of the increased resources available to them. The best results were achieved when placing the processes and threads furthest apart on the node. The graph on the right-hand side of Figure 8 shows the improvement in memory bandwidth that was achieved by using explicit affinity, as reported by the CrayPat performance analysis tool [33].

### D. Implications for Power Consumption

An important side effect of an improvement in runtime is the reduction of power consumption, i.e. the amount of energy that a job consumes to find a solution. Unfortunately there are currently no tools available on our main test systems to be able to quantify exactly how much power is drawn per job, although discussions with hardware vendors confirm that this is an area that is being investigated with increasing interest. Wanting to get a basic understanding of the difference in power consumption between MPI and OpenMP jobs, we used the *likwid-powermeter* tool (part of the likwid toolset [22]) on a single quad-core Intel Core i7 node with hyper-threading[3]. Figure 9 shows the results of the measurements taken for the computation of a CG solve of the Backward Facing Step Velocity matrix, using OpenMP or MPI only. Up to 4 cores, separate physical cores are used; the runs on 8 cores use the processor's hyper-threading capabilities.

---

[3]A single physical core is presented to the OS as two logical cores, however they share the same hardware resources.

| | | | | | |
|---|---|---|---|---|---|
| joules stdev | | | | | |
| MPI Watts | 31.86 | 43.98 | | 47.65 | 48.80 |
| watts stdev | | | | | |

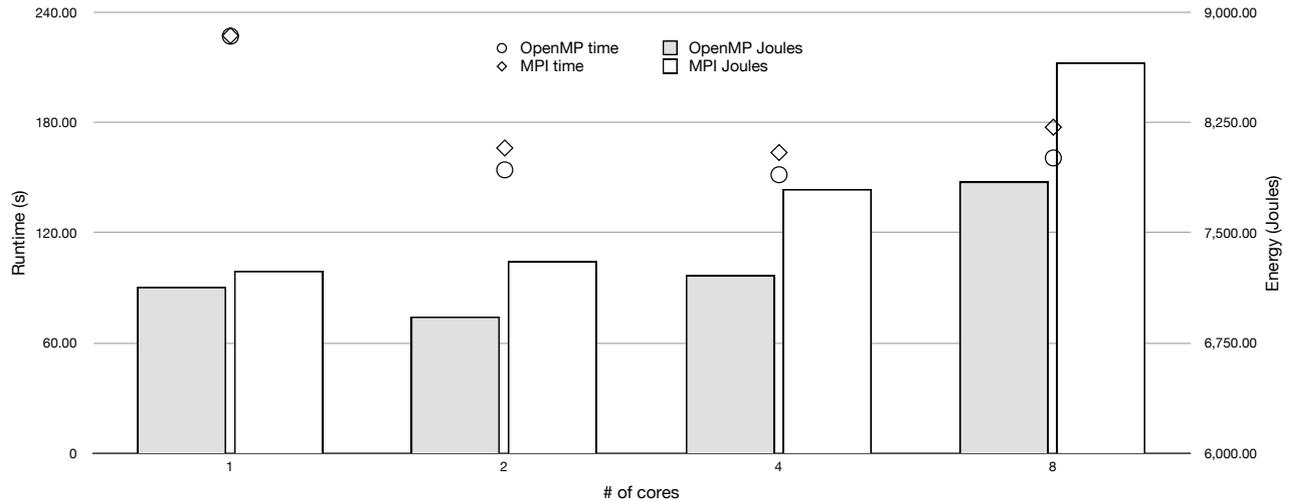

Figure 9: "Energy to solution" for CG solve of the Backward Facing Step Velocity matrix using MPI and OpenMP. These measurements were taken on a quad-core Intel Core i7 node with hyper-threading capability.

The runtimes show that on this particular system the test case does not scale beyond two cores due to limited memory bandwidth. For both OpenMP and MPI, this represents the ideal number of cores for this test: there is no runtime improvement from 2 to 4 cores and thus no performance benefit, however due to the use of extra hardware resources the energy consumption increases. The OpenMP runs use less energy than their MPI counterparts because of their reduced runtimes; in terms of Watts, both programming models exhibit similar behaviour.

The example presented in Figure 9 is a typical case of a memory bandwidth limited execution. There is no performance to be gained from using all the cores on the processor. It is important for programmers and users to be aware of their code's performance profile within a node (and within memory hierarchies): while using all the cores on a processor may not degrade performance (runtimes may simply flatline), it can have a noticeable impact on the amount of energy that a job consumes to no benefit.

E. *Performance across Multiple Nodes*

We have established that the single node performance of the OpenMP implementation of PETSc is outperforming MPI, both in terms of scalability and runtime performance. The next step involves verifying if this superior performance inside a node translates to superior performance across nodes as well. For the hybrid benchmarks, where each process spawns a pool of threads, it is important to place MPI processes equidistantly across the node to guarantee uniform access to the resources; we use *aprun* options to define the distribution of processes and the number of threads per process. On HECToR for example, using 8 threads means that only 4 MPI processes are placed on a 32-core node and each of these processes is placed on its own UMA region.

Figure 10 shows the runtimes we achieved performing a CG solve on the medium-sized Saltfingering pressure matrix. The graph on the left shows the time spent in the KSPSolve function, which represents the complete CG solve including the time taken by the Jacobi pre-conditioner. The second graph is for the time spent in the matrix-vector multiplication only, which is the predominant part of the solver. The benchmark was run on up to 512 cores on fully populated HECToR nodes, comparing MPI with different hybrid configurations. PETSc's implementation of the matrix-vector multiplication distinguishes between on- and off-diagonal elements, which are multiplied in separate functions and later gathered. When reducing the number of MPI processes, the amount of local work per thread increases, but the number of messages needing to be communicated decreases at the same time. Looking at Figure 10, the hybrid performance is nearly always better than the MPI performance. For 8 nodes, the performance is slightly poorer when using more than 2 threads; this is because the penalties incurred for non-local vector accesses are greater than the benefits of using fewer MPI processes. The improved scaling of the hybrid code is most notable on 16 nodes (512 cores), where the MPI code starts slowing down: we reach the point where the amount of work per process is too small and the communication overheads outweigh any performance benefits. The hybrid code however continues to scale, as a reduced number of processors means that threads have more work and fewer messages need to be passed.

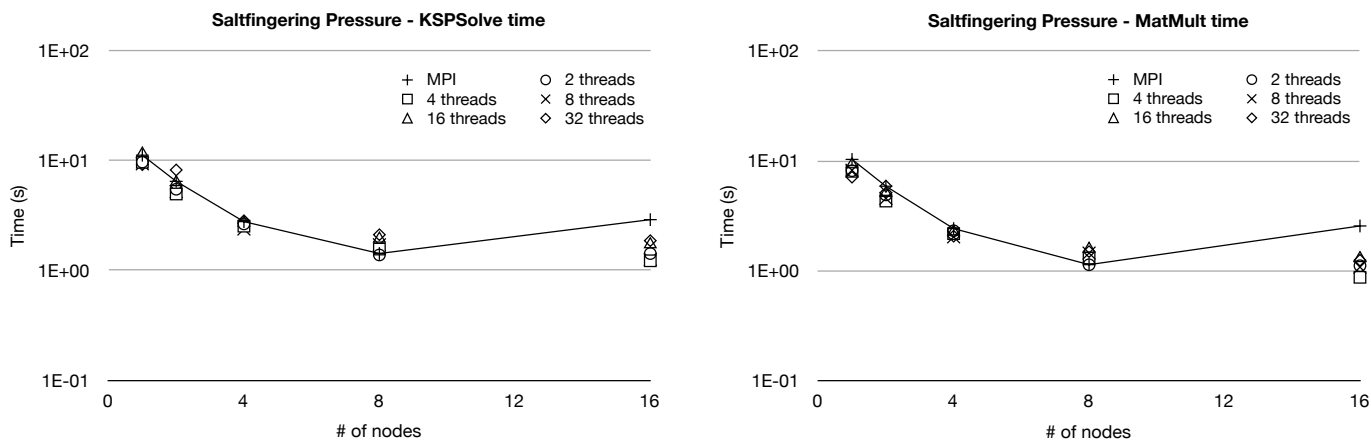

Figure 10: Runtimes for a CG solve of the Saltfingering pressure matrix, with a Jacobi preconditioner. The graph on the left shows the total time spent in the solver function, comparing MPI to hybrid performance from 32 to 512 cores (or 1 to 16 HECToR nodes). On the right is the total time spent in the matrix-vector multiplication, a subset of the KSPSolve function depicted on the left.

Our final test involves the largest matrix, extracted from a flue emission model, with almost 750 million non-zero elements (8.5GB on disk). The size of the matrix allowed us to do larger-scale runs up to 16,384 cores. Figure 11 illustrates the performance improvement of MatMult with shared-memory threading inside an UMA region in comparison with MPI. The performance gains are most dramatic for the highest core counts. Again, for smaller numbers of cores beyond a single node, the benefits of using threads are less pronounced (albeit visible), because the trade-off in performance between using fewer MPI processes and access non-local vector data is less dramatic. For the MPI code strong scaling essentially stops at 2k cores. The hybrid code on the other hand continues to scale, benefiting from the fact that fewer MPI processes are being used, thus reducing the number of messages that are sent and the contention on the network. For 8k cores, our mixed-mode version of PETSc gives a performance improvement of more than 50% for 4 and 8 threads (using 2048 and 1024 MPI processes respectively) over pure MPI. This is a considerable improvement not only of the runtime performance, but also of the efficient usage of the available hardware.

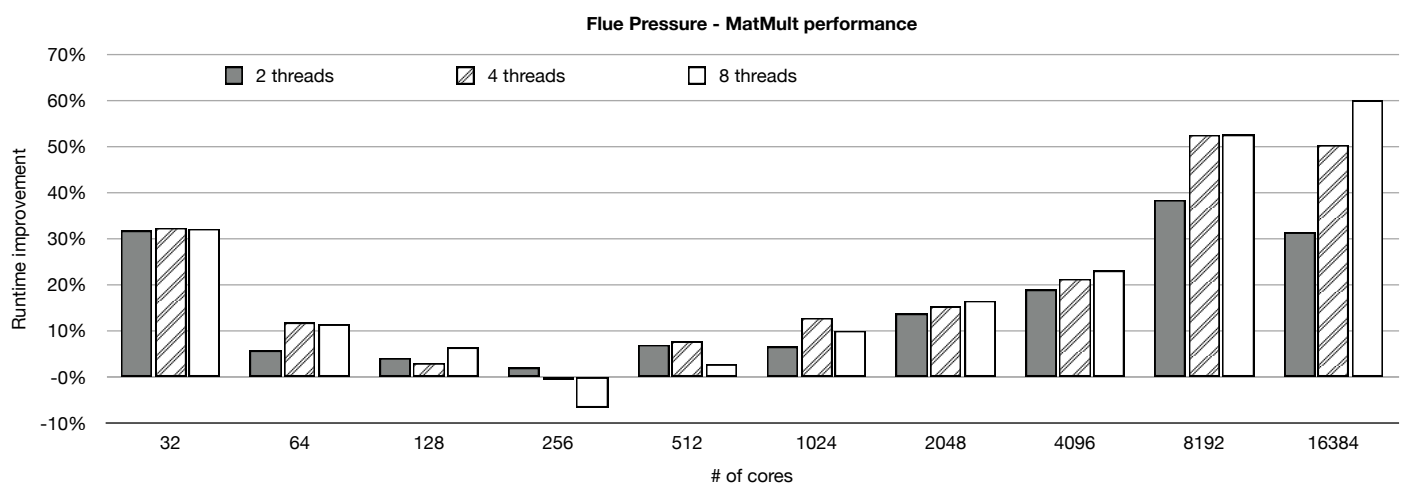

Figure 11: Difference in performance for hybrid MatMult (the sparse matrix-vector multiplication), using threads inside an UMA region, compared to pure MPI, for the MatMult part of a GMRES solve on the Flue matrix. The MPI performance is the baseline (0%).

## IX. CONCLUSIONS AND FUTURE WORK

In order to exploit the power of next-generation exascale systems, scientific applications and libraries need to evolve in step with the hardware. Moving away from the pure message-passing model towards a mixed-mode programming approach, allowing applications to exploit hardware more efficiently, is widely seen as an important step in this evolution. We have added OpenMP support to the PETSc library, widening its target audience to now include mixed-mode applications that use OpenMP alongside MPI, as well as smoothing the path for MPI-only applications that use the library to adopt a hybrid programming model. The implementation of the threaded version ensures that performance-critical decisions, such as data memory paging, are almost entirely handled by the library. The interface to the library functions remains unchanged. Optional threading depending on object size and optimised vector data locality for matrix-vector multiplication are yet to be implemented; both are expected to provide further performance boosts. Despite this, the performance of our current implementation of the hybrid library is already almost always superior to that of the MPI-only implementation.

ACKNOWLEDGMENT

Many thanks to Tom Edwards from Cray UK for sharing his experiences using the AMD Interlagos processor and to the PETSc development team for their support throughout this work.